\documentclass[a4paper,11pt]{article}
\usepackage{pos}
\usepackage{enumerate}

\title{Instanton effects on chiral symmetry breaking and hadron spectroscopy}
\ShortTitle{Instanton effects}

\author*[]{Masayasu Hasegawa}

\affiliation[]{Bogoliubov Laboratory of Theoretical Physics,\\
  Joint Institute for Nuclear Research, Dubna, Moscow 141980, Russia}

\emailAdd{hasegawa@theor.jinr.ru}

\abstract{This project aims to give indications to find monopole and instanton effects in QCD on the observables by experiments. First, we add the monopole and anti-monopole to the QCD vacuum of the quenched SU(3) and calculate the physical observables using the eigenvalues and eigenvectors of the overlap Dirac operator that preserves the exact chiral symmetry. We have found that the additional monopole and anti-monopole make the long monopole loops are closely related to the quark confinement without changing the vacuum structure. Furthermore, we have confirmed that the additional monopole and anti-monopole create instantons and anti-instantons are closely associated with the chiral symmetry breaking. We have shown that the chiral condensate (minus value) decreases in direct proportion to the square root of the number density of the instantons and anti-instantons. The decay constants and masses of pion and kaon increase in direct proportion to the one-fourth root of the number density of the instantons and anti-instantons. This report estimates the eta meson mass using these outcomes as the input values, and the eta-prime meson mass is calculated in two ways: (i) Substituting the numerical results of the topological charge and pion decay constant to the Witten and Veneziano mass formula. (ii) Calculating the correlations of the disconnected (hairpin) graphs. The preliminary results of the eta-prime meson mass estimated in the quenched SU(3) are as follows. (i) m$_{\eta'}$ = 1.055(15)$\times10^{3}$ [MeV] (at the continuum limit). (ii) m$_{\eta'}$ = 1.04(2)$\times10^{3}$ [MeV] (at the chiral and continuum limits). Finally, we demonstrate that the eta-prime meson mass becomes heavy with increasing the number density of the instantons and anti-instantons.}

\FullConference{%
 The 38th International Symposium on Lattice Field Theory, LATTICE2021
  26th-30th July, 2021
  Zoom/Gather@Massachusetts Institute of Technology
}

\begin{document}
\maketitle

\section{Introduction}

Monopoles play critical roles in the quark confinement mechanism through condensing in the QCD vacuum~\cite{tHooft2,Mandelstam1}, and the instantons induce spontaneous chiral symmetry breaking~\cite{Belavi1, Dyakonov6, Shuryak2}. Monopoles and instantons are closely related and interact among quarks and gluons in the QCD vacuum~\cite{Rubakov1,Shuryak2}. It is very interesting if we can show a clue to observe monopoles and instantons by experiments. Therefore, we perform numerical simulations of lattice gauge theory and investigate the effects of monopoles and instantons on hadrons.

First, we apply the monopole creation operator to the QCD vacuum~\cite{Bonati2} and add one pair of monopole and anti-monopole to the vacua of the quenched SU(3), varying the magnetic charges. Second, we estimate the monopoles and instanton effects on observables using the eigenvalues and eigenvectors of the overlap Dirac operator that preserves the exact chiral symmetry~\cite{Ginsparg1, Neuberger1, Neuberger2, Lusher1, Chandrasekharan1}. Finally, we compare the numerical results with the predictions and find quantitative relations among the monopoles, instantons, and observables.

Previous research~\cite{DiGH3} has found that the monopole creation operator makes the monopoles and anti-monopoles. These added monopoles and anti-monopoles form the long monopole loops in the QCD vacuum~\cite{Bode1} that are closely related to color confinement~\cite{Ejiri1}. We have demonstrated that one pair of additional monopole and anti-monopole with magnetic charges creates one instanton or anti-instanton. Furthermore, the additional monopoles and anti-monopoles do not change the vacuum structure.

We compared the numerical results with the predictions and discovered the effects of the added monopole and anti-monopoles and the created instantons and anti-instantons on the observables as follows~\cite{Hasegawa2}:
\begin{enumerate}
\item The added monopole and anti-monopoles do not affect the low-lying eigenvalues of the overlap Dirac operator and only change a scale parameter of the distribution of the low-lying eigenvalues.
  
\item The chiral condensate (defined as a negative value) decreases in direct proportion to the square root of the number density of the instantons and anti-instantons. 
\item The average mass of the quarks ($\frac{u+d}{2}$) and the s-quark mass become heavy in direct proportion to the square root of the number density of the instantons and anti-instantons.
\item The decay constants and masses of the pion and kaon increase in direct proportion to one-fourth root of the number density of the instantons and anti-instantons.
\item The decay width of the charged pion becomes wider than that of the experimental result. As a result, the lifetime of the charged pion becomes shorter than that of the experimental result.
\end{enumerate}     
We obtained these results using two lattices: The lattice volumes $V$ are $V = 14^{4}$ and $V = 18^{3}\times32$, and their values of a parameter $\beta$ for the lattice spacing are $\beta$ = 6.000 and $\beta$ = 6.052, respectively.

This research project investigates the finite lattice volume effect and the discretization effect on these numerical results. Therefore, we generate the various configurations by varying the lattice volumes $V$ and values of the parameter $\beta$.

In this report, the eta-prime meson mass is estimated by the following two calculations: (i) Substituting the numerical results of the pion decay constant and topological susceptibility to the mass formula derived by Witten and Veneziano~\cite{Giusti8}. (ii) Calculating the correlations of the disconnected (hairpin) graphs of the pseudoscalar density~\cite{DeGrand2}. The numerical results of the eta-prime meson mass estimated in the quenched SU(3) are (i) m$_{\eta'}$ = 1.055(15)$\times10^{3}$ [MeV] (at the continuum limit) and (ii) m$_{\eta'}$ = 1.04(2)$\times10^{3}$ [MeV] (at the chiral and continuum limits). These results are reasonably consistent with the experimental result m$_{\eta'}^{exp.}$ = 957.78 $\pm$ 0.06 [MeV]~\cite{PDG_2020}. The preliminary results demonstrate that the eta-prime meson mass becomes heavy with increasing the number density of the instantons and anti-instantons. Now, we are evaluating the increases by comparing them with the predictions. 

The contents of this report are as follows. In section~\ref{sec:1}, we explain the monopole creation operator and simulation parameters very briefly. In section~\ref{sec:2}, we give the results of the number density of the instantons and anti-instantons, chiral condensate, pion decay constant, and masses of pion, kaon, and eta mesons. In section~\ref{sec:2}, we show new results of the mass of the eta-prime meson. Finally, we give the summary and conclusions in section~\ref{sec:3}.

The results of this report are preliminary. We will explain the details of the computations in~\cite{Hasegawa3}.

\section{Monopoles and instantons}\label{sec:1}

This project investigates the finite lattice volume effect and the discretization effect on the numerical results. To check the finite lattice volume effect, we set the parameter $\beta = 6.000$ and vary the lattice volume from $V = 14^{3}\times28$ to $16^{3}\times32$. Similarly, to check the discretization effect, we set the physical lattice volume $V_{phys} = 9.868$ [fm$^{4}$] and vary the lattice volumes and their values of the parameter $\beta$ as follows: $V$ = $12^{3}\times24$, $14^{3}\times28$, $16^{3}\times32$, $18^{3}\times32$, and $20^{3}\times40$, and their $\beta$ = 5.846, 5.926, 6.000, 6.052, and 6.137, respectively.
\begin{table}[htbp]
  \begin{center}
    \caption{The simulation parameters.}\label{tb:parameters}
    \begin{footnotesize}
      \begin{tabular}{|c|c|c|c|c|}\hline
        $\beta$ & $a$ [fm] & $V$ & Conf & $N_{\mbox{conf}}$ \\ \hline
    5.8457 & 0.1242               & $12^{3}\times$24 & Normal conf, $m_{c}=0-4$ & $1.0\times10^{3}\sim 1.2\times10^{3}$ \\\hline
    5.9256 & 0.1065               & $14^{3}\times$28 & Normal conf, $m_{c}=0-5$ & $8\times10^{2}\sim 9\times10^{2}$ \\ \hline
    6.0000 & 9.3150$\times10^{-2}$ & $14^{3}\times$28 & Normal conf, $m_{c}=0-4$ & $1.7\times10^{3}\sim 1.8\times10^{3}$ \\ \cline{3-5}
                 &                                           & $16^{3}\times$32 & Normal conf, $m_{c}=0-5$ & $8\times10^{2}\sim 9\times10^{2}$ \\ \hline
    6.0522 & 8.5274$\times10^{-2}$ & $18^{3}\times$32 & Normal conf, $m_{c}=0-6$ & $8\times10^{2}$ \\ \hline
    6.1366 & 7.4520$\times10^{-2}$ & $20^{3}\times$40 & Normal conf, $m_{c}=4-5$ & $4\times10^{2}$ \\ \hline
      \end{tabular}
    \end{footnotesize}
  \end{center}
\end{table}
\vspace{-3mm}

We generate the normal configurations and configurations to which we add the monopoles and anti-monopoles, varying the magnetic charge $m_{c}$ from 0 to 6. We use the monopole creation operator~\cite{Bonati2,DiGH3}. The monopole has the positive magnetic charge of the integer number, and the anti-monopole has the opposite magnetic charges of the monopole. We add both the monopole and anti-monopole that have the same magnitude of magnetic charges. Thus, the total magnetic charges are zero that are added to the configurations, and the magnetic charge $m_{c}$ indicates that both magnetic charges are added. The simulation parameters are in Table~\ref{tb:parameters}. The values of the lattice spacing are calculated using the formula~\cite{Necco1}.

We then estimate the number of instantons and anti-instantons $N_{I}$ in the configurations using the formula $N_{I} = \langle Q^{2} \rangle$ and the numerical results of the topological charge $Q = n_{+} - n_{-}$ because we have never observed the number of zero-modes of the plus chirality $n_{+}$ and the number of zero-modes of the minus chirality $n_{-}$ at the same time from the same configuration. In this study, the observed zero-modes are the topological charges, and the topological susceptibility $\frac{\langle Q^{2} \rangle}{V}$ is the number density of the instanton and anti-instantons $\frac{N_{I}}{V}$~\cite{DiGH3}.

First, to obtain the value at the continuum limit, we fit the linear function to the data of the number density of the instanton and anti-instantons. However, the fitting results of the slope are almost zero considering their errors. Therefore, we interpolate the number density of the instanton and anti-instantons by the constant function. For the same reason, we interpolate the observables by the constant function. We will show the fitting results in~\cite{Hasegawa3}. The numerical results of $\frac{N_{I}}{V}$ of the configurations $\beta = 6.000$ and interpolated results shown are in Table~\ref{tb:all_data_interp}.

\section{Estimations of $\langle\bar{\psi}\psi\rangle^{\overline{MS}}$, $m_{\pi}$, $m_{k}$, $m_{\eta}$, and $F_{\pi}$}\label{sec:2}

\begin {table}[htbp]
  \begin{center}
    \caption{Numerical results of the number density of the instanton and anti-instantons $\frac{N_{I}}{V}$, renormalized chiral condensate in the $\overline{\mbox{MS}}$-scheme at 2 [GeV] $\langle\bar{\psi}\psi\rangle^{\overline{MS}}$, pion decay constant $F_{\pi}$, masses of pion $m_{\pi}$, kaon $m_{k}$, eta $m_{\eta}$, and eta-prime $m_{\eta'}^{(i)}$ mesons.}\label{tb:all_data_interp}
    \begin{footnotesize}
      \begin{tabular}{|c|c|c|c|c|c|c|c|}\hline
        \multicolumn{8}{|c|}{$V=14^{3}\times28$, $\beta = 6.000$} \\ \hline
        $m_{c}$ & $\frac{N_{I}}{V}$ [GeV$^{4}$] & $\langle\bar{\psi}\psi\rangle^{\overline{MS}}$ [GeV$^{3}$] & $F_{\pi}$ [MeV] & $m_{\pi}$ [MeV] & $m_{k}$ [MeV] & $m_{\eta}$ [MeV] & $m_{\eta'}^{(i)}$ [MeV] \\
        & $\times10^{-3}$ & $\times10^{-2}$ & & $\times10^{2}$ & $\times10^{2}$ & $\times10^{2}$ & $\times10^{3}$ \\ \hline        
        Normal conf & 1.60(6) & -1.96(11) & 92(2)  & 1.40(3) & 4.94(15) & 5.64(17) & 1.06(3)\\ \cline{1-8}
        0 & 1.60(5) & -1.89(10) & 92(2)  & 1.39(3) & 4.88(15) & 5.57(17) & 1.07(3)\\ \cline{1-8}
        1 & 1.74(6) & -2.06(12) & 94(2)  & 1.43(3) & 5.04(16) & 5.76(19) & 1.08(3)\\ \cline{1-8}
        2 & 2.23(7) & -2.27(12) & 99(2)  & 1.50(3) & 5.35(16) & 6.12(18) & 1.16(3)\\ \cline{1-8}
        3 & 2.65(9) & -2.40(13) & 102(2) & 1.54(3) & 5.45(17) & 6.23(2)  & 1.24(3)\\ \cline{1-8}
        4 & 2.77(10)& -2.52(14) & 105(2) & 1.58(3) & 5.60(17) & 6.4(2)   & 1.23(3)\\ \cline{1-8}
        5 & 2.71(9) & -2.54(13) & 105(2) & 1.59(3) & 5.63(16) & 6.44(18) & 1.21(3)\\ \cline{1-8}
        \multicolumn{8}{|c|}{$V=16^{3}\times32$, $\beta = 6.000$} \\ \hline
        Normal conf & 1.54(6) & -1.95(10) & 92.3(1.9) & 1.40(3) & 4.94(15) & 5.64(17) & 1.04(3) \\ \cline{1-8}
        0 & 1.62(8) & -1.96(11) & 92(2)     & 1.40(3) & 4.96(16) & 5.67(19) & 1.07(3) \\ \cline{1-8}
        1 & 1.57(7) & -2.03(12) & 94(2)     & 1.42(3) & 5.07(17) & 5.8(2)   & 1.03(3) \\ \cline{1-8}
        2 & 2.07(10)& -2.23(12) & 99(2)     & 1.49(3) & 5.32(17) & 6.1(2)   & 1.13(4) \\ \cline{1-8}
        3 & 2.17(10)& -2.31(12) & 101(2)    & 1.52(3) & 5.44(18) & 6.2(2)   & 1.13(4) \\ \cline{1-8}
        4 & 2.44(12)& -2.45(13) & 103(2)    & 1.56(3) & 5.50(17) & 6.3(2)   & 1.17(4) \\ \cline{1-8}
        5 & 2.34(11)& -2.44(13) & 103(2)    & 1.56(3) & 5.54(17) & 6.3(2)   & 1.15(4) \\ \cline{1-8}
        \multicolumn{8}{|c|}{Interpolated results} \\ \hline 
        Normal conf & 1.58(3) & -1.95(5) & 92.2(1.0) & 1.396(14) & 4.94(7) & 5.64(9)  & 1.055(15) \\ \cline{1-8}
        0 & 1.64(4) & -1.96(6) & 92.5(1.1) & 1.399(17) & 4.95(9) & 5.66(10) & 1.071(18) \\\cline{1-8}
        1 & 1.75(4) & -2.06(6) & 94.7(1.2) & 1.432(18) & 5.06(9) & 5.78(11) & 1.082(19) \\\cline{1-8}
        2 & 2.08(5) & -2.21(6) & 98.1(1.2) & 1.484(18) & 5.28(10)& 6.04(11) & 1.138(19) \\\cline{1-8}
        3 & 2.25(5) & -2.26(7) & 99.3(1.2) & 1.503(18) & 5.36(10)& 6.13(11) & 1.17(20)  \\\cline{1-8}
        4 & 2.36(6) & -2.32(6) & 100.8(1.1)& 1.525(16) & 5.41(9) & 6.18(10) & 1.181(19) \\\cline{1-8}
        5 & 2.44(7) & -2.39(7) & 102.2(1.1)& 1.546(17) & 5.50(9) & 6.28(11) & 1.18(21) \\\cline{1-8}
      \end{tabular}
    \end{footnotesize}
  \end{center}
\end{table}
\vspace{-3mm}

We calculate the renormalized chiral condensate in the $\overline{\mbox{MS}}$-scheme at 2 [GeV] $\langle\bar{\psi}\psi\rangle^{\overline{MS}}$, pion decay constant $F_{\pi}$, masses of pion $m_{\pi}$, kaon $m_{k}$, and eta-prime $m_{\eta'}$ mesons using the eigenvalues of the massive overlap Dirac operator and eigenvectors of the massless overlap Dirac operator~\cite{DeGrand2}. The definitions of the quark propagator, massless and massive Dirac operator, bare quark masses, notations, etc., are the same as previous research~\cite{Hasegawa3}.
\begin{figure*}[htbp]
  \begin{center}
    \includegraphics[width=150mm]{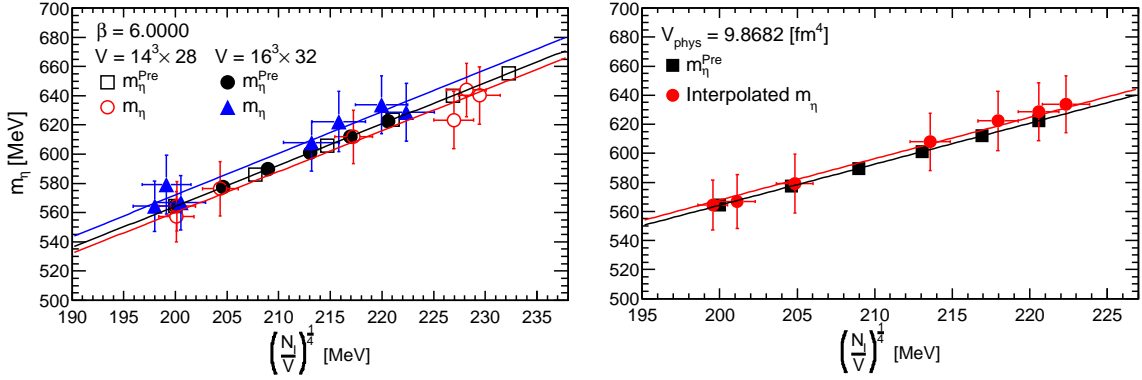}
  \end{center}
  \setlength\abovecaptionskip{-1pt}
  \caption{Estimations of the eta meson mass $m_{\eta}$. The left panel shows the numerical results of $\beta$ = 6.000, $V = 14^{3}\times28$ and $V = 16^{3}\times32$, and the right panel shows the interpolated results. The colored lines represent the fitting results of the numerical results, and the black lines represent the fitting results of the predictions.}\label{fig:meta}
\end{figure*}
\vspace{-3mm}

First, we calculate the correlation of the pseudoscalar density $C_{PS}$ and the correlation of the scalar density $C_{SS}$ using the eigenvalues and eigenvectors and subtract correlations as follows: $C_{PS}-C_{SS}$~\cite{Blum1}. We vary the bare quark mass $m_{q}$ from 30 to 150 [MeV] and evaluate the pseudoscalar mass $m_{PS}$ and decay constant $F_{PS}$ by fitting a curve to the numerical results of the correlation $C_{PS}-C_{SS}$. 

Then, we calculate the intersections by matching the experimental results of the decay constants and masses of pion and kaon and the numerical results of the PCAC relation ($m_{PS}^{2} = Am_{q}$) and determine the normalization factors~\cite{Giusti3}. 

The normalization constant of the scalar $Z_{S}$ is calculated~\cite{Hernandez2,Wennekers1}, and the renormalized chiral condensate in the $\overline{\mbox{MS}}$-scheme at 2 [GeV] $\langle\bar{\psi}\psi\rangle^{\overline{MS}}$ is estimated using the fitting result of the slope $A$ of the PCAC relation. We then estimate the decay constants and masses of pion and kaon using the outcomes of the intersections and normalization factors. The eta meson mass $m_{\eta}$ is estimated from the following mass formula $m_{\eta} = \sqrt{\frac{4m_{k}^{2}}{3} - \frac{m_{\pi}^{2}}{3}}$ using the numerical results of pion and kaon masses. The numerical results and the interpolated results are shown in Table~\ref{tb:all_data_interp}.

To evaluate the increases in the eta meson mass, we fit the curve $m_{\eta'} = A(\frac{N_{I}}{V})^{\frac{1}{4}}$ to the numerical results as shown in Fig~\ref{fig:meta}. The prediction $m_{\eta}^{Pre}$ is calculated using the experimental results of the kaon, pion, and outcome of the phenomenological calculation~\cite{Shuryak2}.

The fitting results are as follows: (1) Prediction; A = 2.822 and $\chi^{2}$/d.o.f. = 0/5. (2) $V = 14^{3}\times28$; A = 2.80(3) and $\chi^{2}$/d.o.f. =  0.6/6.0. (3) $V = 16^{3}\times32$; A = 2.86(4) and $\chi^{2}$/d.o.f. =  0.6/6.0. (4) Interpolated results; A = 2.84(4) and $\chi^{2}$/d.o.f. = 0.1/6.0. The fitting results of the slope $A$ are consistent with the prediction and the values of $\chi^{2}$/d.o.f. are small. Therefore, the eta meson mass becomes heavy in direct proportion to the one-fourth root of the number density of the instanton and anti-instantons.

Similarly, we fit the following curves $\langle\bar{\psi}\psi\rangle^{\overline{MS}} = A(\frac{N_{I}}{V})^{\frac{1}{2}}$, $F_{\pi} = A(\frac{N_{I}}{V})^{\frac{1}{4}}$, and $m = A(\frac{N_{I}}{V})^{\frac{1}{4}}$ to the numerical results. The fitting results are consistent with the predictions. Therefore, the renormalized chiral condensate in the $\overline{\mbox{MS}}$-scheme at 2 [GeV] decreases in direct proportion to the square root of the number density of the instanton and anti-instantons. The pion decay constant increases in direct proportion to the one-fourth root of the number density of the instanton and anti-instantons. The meson masses of pion, kaon, eta become heavy in direct proportion to the one-fourth root of the number density of the instanton and anti-instantons.

\section{Eta-prime meson mass}\label{sec:3}

This last section estimates the eta-prime meson mass by following two computation methods.

(i) We obtain the eta-prime meson mass by substituting the numerical results of the pion decay constant and topological susceptibility, that is, the number density of the instantons and anti-instantons, to the following Witten and Veneziano relation of the leading-order term~\cite{Giusti8}.
  \begin{equation}
    m_{\eta'}^{(i)} = \frac{2N_{f}}{F_{\pi}^{2}}\frac{\langle Q^{2}\rangle}{V} = \frac{2N_{f}}{F_{\pi}^{2}}\frac{N_{I}}{V}, \ (F_{\pi} \approx 94 \ [\mbox{MeV}], \ N_{f} = 3).
  \end{equation}
  
  (ii) The eta-prime meson mass in the quenched approximation $\mu_{0}$ is estimated by calculating the following correlation function $C_{dis-PP}(\Delta t)$ of the disconnected graphs of the pseudoscalar density $\mathcal{O}_{PS}$~\cite{DeGrand2}.
\begin{equation}
  C_{dis-PP}(\Delta t) = \frac{a^{3}}{V} \sum_{t}\langle \sum_{\vec{x}_{2}}\mathcal{O}_{PS}^{C}(\vec{x}_{2}, t) \sum_{\vec{x}_{1}}\mathcal{O}_{PS}(\vec{x}_{1}, t + \Delta t)\rangle\label{eq:cdispp}.
  \end{equation}
The pseudoscalar density is $\mathcal{O}_{PS} = \bar{\psi}_{1}\gamma_{5}\left( 1 - \frac{a}{2\rho}D \right) \psi_{2}$. The correlation function is defined for each flavor. Suppose that this correlation function can be approximated by the following function of the double poles.
\begin{equation}
  C_{dis-PP}(t) = \frac{Z_{PS}}{4m_{PS}}\frac{\mu_{0}^{2}}{N_{f}}\left[(1+m_{PS}t)\exp(-m_{PS}t) + \{1+m_{PS}(T-t)\}\exp\{-m_{PS}(T-t)\}\right]
\end{equation}
We fit this curve to the computed results and obtain the coefficient $Z_{PS}\mu_{0}^{2}$ and $m_{PS}$. We substitute the coefficient $Z_{PS}$ that is obtained by fitting a curve to the calculated results of the correlation function of the connected graph of the pseudoscalar density and then estimate the eta-prime meson mass in the quenched approximation $\mu_{0}$.

The computed results of the correlation function~(\ref{eq:cdispp}) varying bare quark mass from $m_{q}$ = 30 to 150 [MeV] do not show any divergence predicted by the chiral perturbation theory. Therefore, we interpolate the eta-prime meson mass $\mu_{0}$ to the chiral limit $m_{q} \rightarrow 0$ by fitting the linear function. All data points are included in a fitting range, and the fitting results of $\chi^{2}/n.d.f.$ are less than 1. The estimations of $\mu_{0}$ and their fitting results of $\chi^{2}/n.d.f.$ of the normal configurations in Table~\ref{tb:m_etap_or} and the configurations of the additional monopoles and anti-monopoles ($V = 18^{3}\times32$, $\beta = 6.052$) in Table~\ref{tb:m_etap}.
    
Last, we estimate the eta-prime meson mass $m_{\eta'}^{(ii)}$ using the mass formula derived by Witten and Veneziano and the numerical results of $m_{\pi}$ and $m_{k}$ as follows:
\begin{equation}
  m_{\eta'}^{(ii)} = \sqrt{\mu_{0}^{2} + \frac{2m_{k}^{2}}{3} + \frac{m_{\pi}^{2}}{3}}
\end{equation}

First, we estimate the eta-prime meson mass $m_{\eta'}^{(i)}$ using the results of $F_{\pi}$ and $\frac{N_{I}}{V}$ in Table~\ref{tb:all_data_interp}. The analytical results of $m_{\eta'}^{(i)}$ are in the same table.
\begin{figure*}[htbp]
  \begin{center}
    \includegraphics[width=150mm]{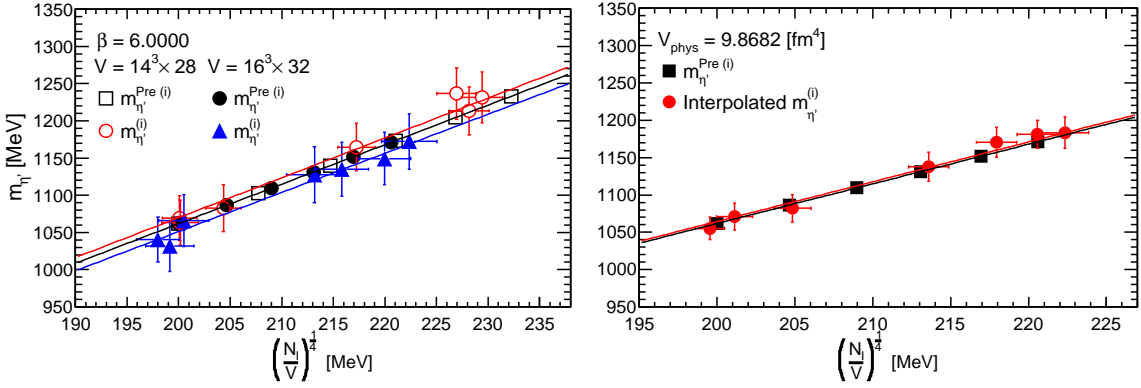}
  \end{center}
  \setlength\abovecaptionskip{-1pt}
  \caption{The estimations of $m_{\eta'}^{(i)}$ compared with the predictions $m_{\eta'}^{Pre (i)}$. The left panel shows the numerical results of $\beta$ = 6.000, $V = 14^{3}\times28$ and $V = 16^{3}\times32$, and the right panel shows the interpolated results. The colored lines represent the fitting results of the numerical results, and the black lines represent the fitting results of the predictions.}\label{fig:meta-prime-1}
\end{figure*}

Figs~\ref{fig:meta-prime-1} show that the eta-prime meson mass $m_{\eta'}^{(i)}$ becomes heavy with increases in the number density of the instantons and anti-instantons $\left(\frac{N_{I}}{V}\right)^{\frac{1}{4}}$. The predictions $m_{\eta'}^{Pre (i)}$ are estimated the experimental results of $m_{\pi}$, $m_{k}$, and the number of instantons that the phenomenological model predicts~\cite{Shuryak2}. To evaluate the rises, we fit the linear function $m_{\eta'} = A(\frac{N_{I}}{V})^{\frac{1}{4}}$, and the fitting results are as follows: (1) Prediction; A = 5.309(18) and $\chi^{2}$/d.o.f. = 0/5. (2) $V = 14^{3}\times28$; A = 5.35(6) and $\chi^{2}$/d.o.f. =  0.6/6.0. (3) $V = 16^{3}\times32$; A = 5.26(7) and $\chi^{2}$/d.o.f. =  0.3/6.0. (4) Interpolated results; A = 5.32(3) and $\chi^{2}$/d.o.f. = 0.7/6.0. These fitting results indicate that the eta-prime meson mass $m_{\eta'}^{(i)}$ increases in direct proportion to the one-fourth root of the number density of the instantons and anti-instantons.
\begin{table}[htbp]
  \begin{center}
    \caption{The results of the normal configurations}\label{tb:m_etap_or}
     \begin{footnotesize}
    \begin{tabular}{|c|c|c|c|}\hline
      $V$ & $\beta$ & $\mu_{0}$ $\times10^{3}$ [MeV] & $\chi^{2}$/n.d.f. \\\hline
      12$^{3}\times$24 & 5.846 & 1.057(16) & 2/19 \\ \cline{1-4}
      14$^{3}\times$28 & 5.926 & 1.057(16) & 1/19 \\ \cline{1-4}
      14$^{3}\times$28 & 6.000 & 1.071(15) & 3/19 \\ \cline{1-4}
      16$^{3}\times$32 & 6.000 & 1.010(15) & 3/19 \\ \cline{1-4}
      18$^{3}\times$32 & 6.052 & 1.034(17) & 3/19 \\ \cline{1-4}
      20$^{3}\times$40 & 6.136 & 0.96(2)   & 3/19 \\ \cline{1-4}
    \end{tabular}
     \end{footnotesize}
  \end{center}
\end{table}
\vspace{-3mm}

Second, we estimate the eta-prime meson mass in the quenched approximation $\mu_{0}$ computed using the standard configurations and interpolate the results to the continuum limit by fitting the linear curve $\mu_{0} = Ax + B$, ($x = a^{2}$, $a$ [fm] is the lattice spacing.). The numerical results $\mu_{0}$ of the standard configurations are shown in Table~\ref{tb:m_etap_or}. The fitting results of the interpolation are $A = 8(2)\times10^{3}$ [MeV]/[fm$^{2}$], $\mu^{0}$ = 9.5(2)$\times10^{2}$ [MeV], and $\chi^{2}$/n.d.f. = 8/3. The eta-prime meson mass of the standard configuration at the continuum limit is $m_{\eta'}^{(ii)}$ = 1.04(2)$\times10^{3}$ [MeV]. Similarly, the eta-prime meson mass $m_{\eta'}^{(ii)}$ of the standard configurations of $\beta$ = 6.000 are $V = 14^{3}\times28$, $m_{\eta'}^{(ii)}$ = 1.147(14)$\times10^{3}$ [MeV] and $V = 16^{3}\times32$, $m_{\eta'}^{(ii)}$ = 1.090(14)$\times10^{3}$ [MeV].
\begin{figure*}[htbp]
  \begin{center}
    \includegraphics[width=95mm]{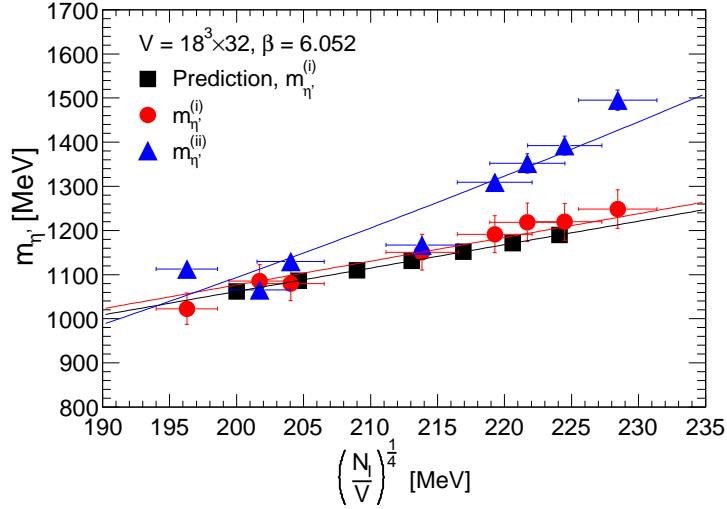}
  \end{center}
  \setlength\abovecaptionskip{-1pt}
  \caption{Comparisons of the eta-prime meson masses $m_{\eta'}^{(i)}$ and $m_{\eta'}^{(ii)}$ with the prediction $m_{\eta'}^{Pre (i)}$. The lattice is V = $18^{3}\times32$, $\beta = 6.052$. The colored lines indicate the fitting results to the numerical results, and the black line indicates the fitting result of the prediction.}\label{fig:meta-prime}
\end{figure*}
\vspace{-3mm}
\begin{table}[htbp]
  \begin{center}
    \caption{The numerical results of the eta-prime meson mass $m_{\eta'}^{(i)}$, $\mu_{0}$, and $m_{\eta'}^{(ii)}$ comparing with the prediction $m_{\eta'}^{Pre}$. The lattice is $V = 18^{3}\times32, \beta = 6.052$.}\label{tb:m_etap}
    \begin{footnotesize}
      \begin{tabular}{|c|c|c|c|c|c|}\hline        
        $m_{c}$ & $m_{\eta'}^{Pre}$ $\times10^{3}$ [MeV] & $m_{\eta'}^{(i)}$ $\times10^{3}$ [MeV] & $\mu_{0}$ $\times10^{3}$ [MeV] & $m_{\eta'}^{(ii)}$ $\times10^{3}$ [MeV] & $\chi^{2}$/n.d.f. \\\hline
      Normal conf & 1.0618(10) & 1.02(4) & 1.034(17) & 1.113(16) & 3/19 \\ \cline{1-6}
      0 & 1.0618(10) & 1.09(4) & 0.984(16) & 1.066(15) & 7/19 \\\cline{1-6}
      1 & 1.0864(10) & 1.08(4) & 1.048(17) & 1.130(17) & 4/19 \\\cline{1-6}
      2 & 1.1095(10) & 1.15(4) & 1.082(19) & 1.167(18) & 7/19 \\\cline{1-6}
      3 & 1.1312(10) & 1.19(4) & 1.23(2)   & 1.31(2)   & 7/19 \\\cline{1-6}
      4 & 1.1517(11) & 1.22(4) & 1.27(2)   & 1.35(2)   & 9/19 \\\cline{1-6}
      5 & 1.1712(11) & 1.22(4) & 1.32(2)   & 1.39(2)   & 3/19 \\\cline{1-6}
      6 & 1.1897(11) & 1.25(4) & 1.42(2)   & 1.49(2)   & 6/19 \\\cline{1-6}
      \end{tabular}
    \end{footnotesize}
  \end{center}
\end{table}
\vspace{-3mm}

Last, we compare the outcomes of $m_{\eta'}^{(i)}$ and $m_{\eta'}^{(ii)}$ computed using the standard configurations and configurations that the monopoles and anti-monopoles are added with the predictions as shown in Fig~\ref{fig:meta-prime}. The lattice is V = $18^{3}\times32$, $\beta = 6.052$, and the computed results are indicated in Table~\ref{tb:m_etap}.

We fit two curves $m_{\eta'}^{(i)} = A^{(i)}(\frac{N_{I}}{V})^{\frac{1}{4}}$ for $m_{\eta'}^{(i)}$ and $m_{\eta'}^{(ii)} = A^{(ii)}(\frac{N_{I}}{V})^{\frac{1}{2}}$ for $m_{\eta'}^{(ii)}$. The fitting results are as follows: (1) $A^{(i)}$ = 5.38(7) and $\chi^{2}$/d.o.f. = 2/7. (2) $A^{(ii)}$ = $2.73(3)\times10^{-2}$ [MeV$^{-1}$] and $\chi^{2}$/d.o.f. = 14/7. The fitting results indicate that the eta-prime meson mass $m_{\eta'}^{(ii)}$ would become heavy in direct proportion to the square root of the number density of the instantons and anti-instantons. 

\section{Summary and conclusions}\label{sec:4}

We calculated the instanton density, evaluated renormalized chiral condensate in the $\overline{\mbox{MS}}$-scheme at 2 [GeV], and estimated the pion decay constant and masses of pion, kaon, eta, and eta-prime mesons using the various configurations. In addition, we investigated the finite lattice volume effect and the discretization effect on these observables and evaluated the numerical results at the continuum limit by interpolation.

We have confirmed that the chiral condensate decreases in direct proportion to the square root of the number density of the instantons and anti-instantons. The pion decay constant and masses of pion, kaon, and eta mesons increase in direct proportion to the one-fourth root of the number density of the instantons and anti-instantons.

Two computations estimated the eta-prime meson mass. First, the eta-prime meson mass $m_{\eta'}^{(i)}$ becomes heavy in direct proportion to the one-fourth root of the number density of the instantons and anti-instantons. However, the eta-prime meson mass $m_{\eta'}^{(ii)}$ becomes heavy in direct proportion to the square root of the number density of the instantons and anti-instantons. Now we are investigating the reason for the difference of increases.

\acknowledgments
We use the SX-series, computer clusters, and XC40 at the Research Center for Nuclear Physics and the Cybermedia Center at Osaka University and the Yukawa Institute for Theoretical Physics at Kyoto University. We used storage elements from the Japan Lattice Data Grid at the Research Center for Nuclear Physics at Osaka University. We appreciate the technical support and computer resources provided by these facilities.

\bibliographystyle{unsrt}

\bibliography{Chiral_symmetry_breaking_and_catalytic_effect_30092020}

\begin{thebibliography}{10}

\bibitem{tHooft2}
{G. 't Hooft}.
\newblock in Proceedings of the EPS International, edited by A. Zichichi, p.
  1225, (1976).

\bibitem{Mandelstam1}
{S. Mandelstam}.
\newblock {II. Vortices and quark confinement in non-Abelian gauge theories}.
\newblock {\em Phys. Rep.}, 23:245, 1976.

\bibitem{Belavi1}
{A. A. Belavin, A. M. Polyakov, A. S. Schwartz, and Yu. S. Tyupkin}.
\newblock {Pseudoparticle solutions of the Yang-Mills equations}.
\newblock {\em Phys. Lett.}, B59:85, 1975.

\bibitem{Dyakonov6}
D.~Diakonov.
\newblock Instantons at work.
\newblock {\em Prog. Particle and Nuclear Physics}, 51:173, 2003.

\bibitem{Shuryak2}
{T. Sch\"{a}fer and E. V. Shuryak}.
\newblock Instantons in {QCD}.
\newblock {\em Rev. Mod. Phys.}, 70(2):323, 1998.

\bibitem{Rubakov1}
{V. A. Rubakov}.
\newblock {Superheavy magnetic monopoles and decay of the proton}.
\newblock {\em Pis'ma Zh. Eksp. Teor. Fiz.}, 33:658, 1981.

\bibitem{Bonati2}
{C. Bonati, G. Cossu, M. D'Elia, and A. Di Giacomo}.
\newblock {The disorder parameter of dual superconductivity in QCD revisited}.
\newblock {\em Phys. Rev. D}, 85:065001, 2012.

\bibitem{Ginsparg1}
{P. H. Ginsparg and K. G. Wilson}.
\newblock {A remnant of chiral symmetry on the lattice}.
\newblock {\em Phys. Rev. D}, 25:2649, 1982.

\bibitem{Neuberger1}
H.~Neuberger.
\newblock Exactly massless quarks on the lattice.
\newblock {\em Phys. Lett. B}, 417:141, 1998.

\bibitem{Neuberger2}
H.~Neuberger.
\newblock More about exactly massless quarks on the lattice.
\newblock {\em Phys. Lett. B}, 427:353, 1998.

\bibitem{Lusher1}
{M. L\"{u}scher}.
\newblock {Exact chiral symmetry on the lattice and the Ginsparg-Wilson
  relation}.
\newblock {\em Phys. Lett.}, B428:342, 1998.

\bibitem{Chandrasekharan1}
{S. Chandrasekharan}.
\newblock {Lattice QCD with Ginsparg-Wilson fermions}.
\newblock {\em Phys. Rev. D}, 60:074503, 1999.

\bibitem{DiGH3}
A.~Di Giacomo and M.~Hasegawa.
\newblock Instantons and monopoles.
\newblock {\em Phys. Rev. D}, 91:054512, 2015.

\bibitem{Bode1}
{A. Bode, T. Lippert, and K. Schilling}.
\newblock {Monopole clusters and critical dynamics in four-dimensional U(1)}.
\newblock {\em Nucl. Phys. B, Proc. Suppl.}, 34:549, 1994.

\bibitem{Ejiri1}
{S. Ejiri, S. Kitahara, Y. Matsubara, and T. Suzuki}.
\newblock {String tension and monopoles in T $\neq$ 0 SU(2) QCD}.
\newblock {\em Phys. Lett.}, B343:304, 1995.

\bibitem{Hasegawa2}
M.~Hasegawa.
\newblock {Monopole and instanton effects in QCD}.
\newblock {\em J. High Energy Phys.}, 09(113), 2020.

\bibitem{Giusti8}
{L. Giusti, G.C. Rossi, M. Testa, G. Veneziano}.
\newblock {The U$_{\mbox{A}}$(1) problem on the lattice with Ginsparg–Wilson
  fermions}.
\newblock {\em Nucl. Phys. B}, 628:234, 2002.

\bibitem{DeGrand2}
{Thomas DeGrand and Urs M. Heller}.
\newblock {Witten-Veneziano relation, quenched QCD, and overlap fermions}.
\newblock {\em Phys. Rev. D}, 65:114501, 2002.

\bibitem{PDG_2020}
{P.A. Zyla et al., (Particle Data Group)}.
\newblock {Review of Particle Physics}.
\newblock {\em Prog. Theor. Exp. Phys.}, 083C01, 2020.

\bibitem{Hasegawa3}
M.~Hasegawa.
\newblock {Chiral symmetry breaking and catalytic effect induced by monopole
  and instanton creations in QCD, in preparation}.

\bibitem{Necco1}
S.~Necco and R.~Sommer.
\newblock {The Nf=0 heavy quark potential from short to intermediate
  distances}.
\newblock {\em Nucl. Phys. B}, 622:328, 2002.

\bibitem{Blum1}
{T. Blum, P. Chen, N. Christ, C. Cristian, C. Dawson, G. Fleming, A. Kaehler,
  X. Liao, G. Liu, C. Malureanu, R. Mawhinney, S. Ohta, G. Siegert, A. Soni, C.
  Sui, P. Vranas, M. Wingate, L. Wu, and Y. Zhestkov}.
\newblock {Quenched lattice QCD with domain wall fermions and the chiral
  limit}.
\newblock {\em Phys. Rev. D}, 69:074502, 2004.

\bibitem{Giusti3}
{L. Giusti, C. Hoelbling, and C. Rebbi}.
\newblock Light quark masses with overlap fermions in quenched {QCD}.
\newblock {\em Phys. Rev. D}, 64:114508, 2001.
\newblock {E}rratum, Phys. Rev. D {\bf 65}, 079903(E) (2002).

\bibitem{Hernandez2}
{P. Hern\'{a}ndez, K. Jansen, L. Lellouch, and H. Wittig}.
\newblock {Non-perturbative renormalization of the quark condensate in
  Ginsparg-Wilson regularizations}.
\newblock {\em J. High Energy Phys.}, 07:018, 2001.

\bibitem{Wennekers1}
J.~Wennekers and H.~Wittig.
\newblock On the renormalized scalar density in quenched {QCD}.
\newblock {\em J. High Energy Phys.}, 09:059, 2005.

\end{thebibliography}

\end{document}